\documentclass[sigconf]{acmart}

\pdfoutput=1

\usepackage{booktabs}
\usepackage{multirow}
\usepackage[nomain, acronym, section]{glossaries} 
\glsdisablehyper
\usepackage{pifont} % for check mark symbol
\usepackage{pgfplots}
\usepackage[utf8]{inputenc}
\pgfplotsset{compat=1.12}
\usepackage{tikz}
\usetikzlibrary{arrows}
\usetikzlibrary{arrows.meta}
\usetikzlibrary{fit}
\usetikzlibrary{positioning}
\usetikzlibrary{shapes.geometric}
\usetikzlibrary{decorations.pathreplacing}
\usetikzlibrary{patterns}
\usetikzlibrary{shapes}
\usetikzlibrary{shadows}
\usepackage[xlatin,abbreviations]{foreign}

\usepackage{etoolbox}
\apptocmd{\sloppy}{\hbadness 10000\relax}{}{}

\usepackage{enumitem}

\newacronym{BIOS}{BIOS}{Basic Input/Output System}
\newacronym{UEFI}{UEFI}{Unified Extensible Firmware Interface}
\newacronym{TCG}{TCG}{Trusted Computing Group}
\newacronym{TPM}{TPM}{Trusted Platform Module}
\newacronym{MBR}{MBR}{Master Boot Record}
\newacronym{ROM}{ROM}{Read-Only Memory}
\newacronym{PCI}{PCI}{Peripheral Component Interconnect}
\newacronym{ACPI}{ACPI}{Advanced Configuration and Power Interface}
\newacronym[shortplural={SCRTMs}]{SCRTM}{SCRTM}{Static Core Root of Trust for Measurement}
\newacronym{SRTM}{SRTM}{Static Root of Trust for Measurement}
\newacronym{CFG}{CFG}{Control-Flow Graph}
\newacronym[shortplural={PCRs}]{PCR}{PCR}{Platform Configuration Register}
\newacronym{DRTM}{DRTM}{Dynamic Root of Trust for Measurement}
\newacronym{PI}{PI}{Platform Initialization}
\newacronym{SEC}{SEC}{Security phase}
\newacronym{PEI}{PEI}{Pre-EFI Initialization phase}
\newacronym{DXE}{DXE}{Driver Execution Environment phase}
\newacronym{BDS}{BDS}{Boot Device Selection phase}
\newacronym{TLS}{TLS}{Transient System Load phase}
\newacronym{RT}{RT}{Runtime phase}
\newacronym{AL}{AL}{After Life phase}
\newacronym{OS}{OS}{Operating System}
\newacronym{DFG}{DFG}{Data Flow Graph}

\newcommand{\tikzcircle}[2][black,fill=black]{\tikz[baseline=-0.5ex]\draw[#1,radius=#2] (0,0) circle ;}%
\newcommand{\xmark}{\tikzcircle[black, fill=white]{2.5pt}}
\newcommand{\cmark}{\tikzcircle{2.5pt}}
\newcommand{\emark}{\ding{55}}

\newenvironment{customlegend}[1][]{%
    \begingroup
    \csname pgfplots@init@cleared@structures\endcsname
    \pgfplotsset{#1}%
}{%
  \csname pgfplots@createlegend\endcsname
    \endgroup
}%

\def\addlegendimage{\csname pgfplots@addlegendimage\endcsname}

\newcommand{\specialcell}[2][c]{\begin{tabular}[#1]{@{}c@{}}#2\end{tabular}}

\newcommand{\EDKII}{EDK~II}
\newcommand{\SeaBIOS}{SeaBIOS}

\begin{document}

\copyrightyear{2019}
\acmYear{2019}
\setcopyright{acmcopyright}
\acmConference[CODASPY '19]{Ninth ACM Conference on Data and Application
Security and Privacy}{March 25--27, 2019}{Richardson, TX, USA}
\acmBooktitle{Ninth ACM Conference on Data and Application Security and Privacy
(CODASPY '19), March 25--27, 2019, Richardson, TX, USA}
\acmPrice{15.00}
\acmDOI{10.1145/3292006.3300026}
\acmISBN{978-1-4503-6099-9/19/03}

\settopmatter{printacmref=true}
\fancyhead{}

\title{BootKeeper: Validating Software Integrity Properties on Boot Firmware Images}

\author{Ronny Chevalier}
\affiliation{\institution{CentraleSup\'elec, Inria, Univ Rennes, CNRS, IRISA}}
\email{ronny.chevalier@centralesupelec.fr}

\author{Stefano Cristalli}
\affiliation{\institution{Universit\`a degli Studi di Milano}}
\email{stefano.cristalli@studenti.unimi.it}

\author{Christophe Hauser}
\affiliation{\institution{University of Southern California}}
\email{hauser@isi.edu}

\author{Yan Shoshitaishvili}
\affiliation{\institution{Arizona State University}}
\email{yans@asu.edu}

\author{Ruoyu Wang}
\affiliation{\institution{Arizona State University}}
\email{fishw@asu.edu}

\author{Christopher Kruegel}
\affiliation{\institution{University of California, Santa Barbara}}
\email{chris@cs.ucsb.edu}

\author{Giovanni Vigna}
\affiliation{\institution{University of California, Santa Barbara}}
\email{vigna@cs.ucsb.edu}

\author{Danilo Bruschi}
\affiliation{\institution{Universit\`a degli Studi di Milano}}
\email{bruschi@di.unimi.it}

\author{Andrea Lanzi}
\affiliation{\institution{Universit\`a degli Studi di Milano}}
\email{andrea.lanzi@unimi.it}

% The default list of authors is too long for headers}
\renewcommand{\shortauthors}{Chevalier \etal}

\begin{abstract} 
    Boot firmware, like \acrshort{UEFI}-compliant firmware, has been the target of numerous attacks,
    giving the attacker control over the entire system while being undetected.
    The \emph{measured boot} mechanism of a computer platform ensures its integrity by using cryptographic measurements to detect such attacks.
    This is typically performed by relying on a \gls{TPM}.
    Recent work, however, shows that vendors do not respect the specifications that have been devised to ensure
    the integrity of the firmware's loading process.
    As a result, attackers may bypass such measurement mechanisms and successfully load a modified firmware image while
    remaining unnoticed.
    In this paper we introduce BootKeeper, a static analysis approach verifying a set of key security properties on
    boot firmware images before deployment,
    to ensure the integrity of the measured boot process.
    We evaluate BootKeeper against several attacks on common boot firmware implementations and demonstrate
    its applicability.
\end{abstract}

%
% The code below should be generated by the tool at
% http://dl.acm.org/ccs.cfm
% Please copy and paste the code instead of the example below. 
%



\maketitle

\section{Introduction}
One of the most critical components of every computer is the boot
firmware (\eg BIOS or UEFI-compliant firmware), which is in charge of
initializing and testing the various hardware components, and then
transfer execution to the \gls{OS}. As a result of its early
execution, the boot firmware is a highly privileged program. Any
malicious alteration of its behavior can have critical consequences on
the entire system. An attacker that can control the firmware can
control any parts of the software and undermine the security of the
entire \gls{OS}. Without any protection of the integrity of the boot
firmware, we cannot assure any security properties of the software
executing on the system.

To guarantee the software integrity of the machine, the \gls{TCG}, an
industry coalition formed to implement trusted computing concepts
across personal computers, designed a new set of hardware components,
the aim of which is to solve various hardware-level trust issues.  In
their specification, they define the \gls{TPM}, which is composed of a
co-processor that offers cryptographic functions (\eg SHA-1, RSA,
random number generator, or HMAC) and a tamper-resistant non-volatile
memory used for storing cryptographic keys~\cite{trusted2007tpm}. The
\gls{TPM} along with other software components together form a root of
trust, which is leveraged as part of several security mechanisms,
including the \emph{measured boot} process.  With measured boot,
platforms with a \gls{TPM} can be configured to measure every
component of the boot process, including the firmware, boot loader,
and kernel.  Such measurement process is also called the \gls{SRTM}.

The core of trust of the entire process is established based on the
integrity of the first piece of code inside the boot firmware which is
doing the first measurements, also called the
\gls{SCRTM}~\cite{zimmer2009trusted}. In the event where a malicious modification
of the \gls{SCRTM} successfully hides from the self-measurement
technique, the whole chain of trust and, consequently, the
integrity of the entire system may be broken.
In this direction, \citet{butterworth2013bios}
described different examples of attacks against the \gls{SCRTM}
component.
In particular, the authors show how a novel ``tick'' malware, a
51-byte patch to the \gls{SCRTM}, can replay a forged measurement to the \gls{TPM},
falsely indicating that the BIOS is genuine.
These attacks take
advantage of the fact that some vendors do not measure the
\gls{SCRTM} code, thus allowing an attacker to modify it and to forge
measurements without being detected.

Recent platforms incorporate an immutable, hardware protected 
\gls{SCRTM}~\cite{intelBootGuard,hpSureStart}.
Intel Boot Guard and HP Sure
Start are immutable \glspl{SCRTM} which measure and verify, at boot time, the integrity of the BIOS image before executing it.
Such technologies are not directly vulnerable to the aforementioned attacks,
since their code cannot be modified by an attacker.\footnote{Nonetheless, Intel Boot Guard has been shown to be vulnerable to some attacks as well~\cite{alex}.}
Both technologies, however, are only available in recent Intel and HP platforms,
\emph{leaving previous hardware implementations, or devices of other vendors,
vulnerable against forged measurements}.
In such implementations, since the \gls{SCRTM} is not hardware protected, \emph{it is usually attached to the firmware image itself during the firmware update process}.
Even when the firmware image is signed, attacks may compromise this process~\cite{kim2017}, and consequently allow an attacker to modify both the firmware code and the \gls{SCRTM}.

In order to solve these challenges in validating the \gls{SCRTM} code, we design a self-contained approach based on static analysis at the binary level, which is able, starting from a boot firmware image, to validate the correctness of the measurement process. 
Our system verifies software properties on the \gls{SCRTM} code embedded in firmware images, including: (1) the completeness of firmware code measurements in terms of fingerprinted 
memory regions, (2) the correctness of cryptographic functions
implemented\footnote{Vendors typically implement the cryptographic functionalities used as part of the measurement process in software.} inside the \gls{SCRTM} (\eg SHA-1), and (3) the
correctness of the \gls{SCRTM}'s control flow.
More in detail, the first property ensures that the code of the entire firmware is measured correctly by the \gls{SCRTM}, \ie that none of the instructions to be executed at runtime will be missed by the measurement process. The second property ensures that the implementation of cryptographic functions inside the \gls{SCRTM} is correct. The third property validates the correctness of the measurement operations performed by the \gls{SCRTM} in terms of execution order. It also guarantees the atomicity of operations occurring between memory fingerprinting and write operations performed on the \gls{TPM} component (\ie ensuring that what is measured is what is written to the TPM). 
Altogether, this set of properties can prevent attacks aiming to elude the measurement process,
and it guarantees that the integrity of a firmware image is properly verified during the \emph{measured boot} process.

We implement a prototype of our system, dubbed BootKeeper, based on the angr program analysis
framework~\cite{angr,shoshitaishvili2016state}.
We evaluate our system on different open source boot firmware images,
and we implement different attacks against the firmware to
show the efficacy of our approach.
Our paper makes the following contributions:
\begin{itemize}
    \item We devise a set of software properties that can be used for validating the measurement process
          and mitigate firmware attacks aimed to subvert the entire system.

    \item We design and implement BootKeeper, a binary analysis approach to detect and prevent measurement boot
          firmware attacks in different attack scenarios.

    \item We perform experimental evaluation against different attacks and several boot firmware implementations
          to demonstrate the effectiveness of our approach.
\end{itemize}

%%% Local Variables: 
%%% mode: latex
%%% TeX-master: "../paper"
%%% End: 

\section{Background}
In this section, we introduce the background technology needed to
understand our approach.
We first describe the principles of the \gls{TPM}, then we describe the UEFI specifications and some of the
software/hardware components involved in the boot measurement process.

\subsection{Trusted Platform Module (TPM)}
The \gls{TPM} specification defines a co-processor offering
cryptographic features (\eg SHA-1, RSA, random number generator, or
HMAC), and tamper-resistant storage for cryptographic
keys~\cite{trusted2007tpm}.

The \gls{TPM} provides a minimum of 16 \gls{PCR} which are
160-bit wide registers used to store the measurements done by the
\gls{SCRTM} (usually SHA-1 hashes). The design of these registers
allows an unlimited amount of measurements and prevents an attacker from
overwriting them with arbitrary values. In order to do this, the only
possible operation is \textit{extend}:
\[
PCR_i = H(PCR_i\ ||\ m)
\]

Where $PCR_i$ is the $i$th \gls{PCR} register, $H$ is the hash function and $m$ the new measurement.
The \gls{TPM} \emph{concatenates} each new measurement sent with the previous value of the register, then it hashes the
result, which becomes the new value of the register.
This mechanism is crucial to establishing a chain of trust, since the only way to obtain a given measurement value
from a \gls{PCR} is to reproduce the same series of measurements in the same order.
The measured boot process relies on this mechanism to guarantee that a given software platform is valid and has not
been tampered with.

The \gls{TPM} also relies on these measurements to provide specific features
(\eg secure storage or remote attestation).
For instance, with the sealing operation, the \gls{TPM} offers the
ability to encrypt data, with a key only known to the \gls{TPM},
and it binds the decryption to the \glspl{PCR} values.
During the decryption (\ie unsealing),
the \gls{TPM} only decrypts the data if the \glspl{PCR}
values match the ones used during the encryption.
One common use case for the sealing operation is to store
the disk encryption key.
It ensures that the disk is decrypted only if the platform
has booted with the expected hardware and software,
and if no attacker tampered with the boot process
(\eg an evil maid attack).

\subsection{Static Core Root of Trust for Measurement}
The \gls{SCRTM} is responsible for the first measurement sent to
the \gls{TPM} in the PCR0 register and it is considered trusted by default
on the system.
Since the default 
values of the \gls{PCR}s are known (either \verb~0x00...0x00~ or 
\verb~0xFF...0xFF~), the entire trust in the \gls{SRTM} relies on the 
\gls{SCRTM}. If it is possible for an attacker to modify the \gls{SCRTM}, then it is also possible for the attacker to forge the first measurement, and the next one, etc.

Therefore, the \gls{TCG} PC client specific
implementation~\cite{tcg2005client} states that the \gls{SCRTM}
must be an immutable portion of the firmware. The specification
defines immutability such that only an approved agent and method can
modify the \gls{SCRTM}. Most firmware fulfill this
requirement by using signed updates, because the \gls{SCRTM}
can only be modified if the update is coming from the vendor.
Recent firmware fulfill the requirement using an immutable
hardware protected \gls{SCRTM}~\cite{intelBootGuard, hpSureStart}.
Unfortunately, legacy platforms do not provide signed updates, or do
not require them. Furthermore, \citet{kallenberg2013defeating}, and \citet{wojtczuk2009attacking} have successfully exploited
multiple vulnerabilities in the implementation of signed firmware
updates by vendors, allowing an attacker to update the firmware with
a malicious one. Finally, if the private key of the vendor is
compromised, the platform is vulnerable.

\subsection{Unified Extensible Firmware Interface}
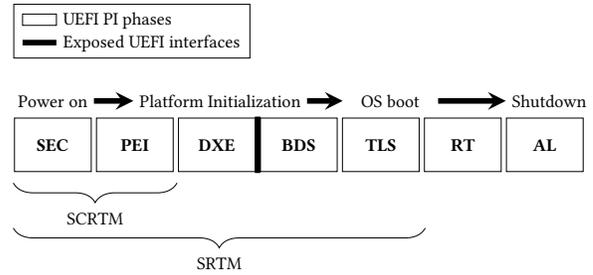
\begin{figure}[ht]
    \centering
    \scalebox{0.75}{
            \begin{tikzpicture}[%
        node distance=0.25em,
        phase/.style={%
            rectangle,
            minimum width=4.3em,
            minimum height=3em,
            draw
        }
    ]
        \node[phase] (1)              {\textbf{\acrshort{SEC}}};
        \node[phase] (2) [right=of 1] {\textbf{\acrshort{PEI}}};
        \node[phase] (3) [right=of 2] {\textbf{\acrshort{DXE}}};
        \node[phase] (4) [right=of 3] {\textbf{\acrshort{BDS}}};
        \node[phase] (5) [right=of 4] {\textbf{\acrshort{TLS}}};
        \node[phase] (6) [right=of 5] {\textbf{\acrshort{RT}}};
        \node[phase] (7) [right=of 6] {\textbf{\acrshort{AL}}};

        \node (POWERON)  [above=of 1] {Power on};
        \node (PINIT)    [right=0.7cm of POWERON] {Platform Initialization};
        \node (OSBOOT)   [above=of 5,right=2em of PINIT,minimum width=5.3em] {\acrshort{OS} boot};
        \node (SHUTDOWN) [above=of 7,right=1.2cm of OSBOOT] {Shutdown};

        \draw [draw=black,solid,line width=1mm] (3.63,-0.5) -- (3.63,0.5);

        \path [draw=black,solid,line width=1mm,->,>=stealth]
            (POWERON) edge node {} (PINIT)
            (PINIT) edge node {} (OSBOOT)
            (OSBOOT) edge node {} (SHUTDOWN);

        \draw [decorate,decoration={brace,mirror,amplitude=10pt},yshift=-5pt]
            (-0.7,-0.5) -- (2.2,-0.5)
            node [black,midway,yshift=-2em]
            {\acrshort{SCRTM}};

        \draw [decorate,decoration={brace,mirror,amplitude=10pt},yshift=-5pt]
            (-0.7,-1.3) -- (6.6,-1.3)
            node [black,midway,yshift=-2em]
            {\acrshort{SRTM}};

        \begin{customlegend}[legend cell align=left,
        legend style={at={(3.5,2.5)}},
        legend entries={
        \acrshort{UEFI} \acrshort{PI} phases,
        Exposed \acrshort{UEFI} interfaces
        }]
            \addlegendimage{area legend}
            \addlegendimage{draw=black,solid,line width=1mm}
        \end{customlegend}
    \end{tikzpicture}
    }
    \Description{A figure describing the several UEFI PI phases where the SCRTM encapsulates the SEC
    and PEI phases, while the SRTM encapsulates the SEC, PEI, DXE, and BDS phases.}
    \caption{\gls{UEFI} \gls{PI} phases with the ones corresponding to the \gls{SCRTM} and \gls{SRTM}}
    \label{fig:uefi_phases}
\end{figure}

In 2005, 11 industry leading technology companies created the
\gls{UEFI} forum which defines specifications for interfaces~\cite{uefi2017}
used by the \gls{OS} to communicate with the
firmware~\cite{zimmer2010beyond} and \gls{PI}
specifications~\cite{uefi2015PI} which define the required interfaces
for the components in the firmware, allowing multiple providers to
create different parts.
\gls{UEFI} specifications are about the interfaces, while
\gls{UEFI} \gls{PI} specifications are about building \gls{UEFI}-compliant firmware.
Manufacturers are now providing, as boot
firmware replacing the \gls{BIOS}, \gls{UEFI}-compliant
firmware images.
\gls{UEFI} and \gls{PI} specifications define seven phases, 
as illustrated in~\autoref{fig:uefi_phases}, which describe the boot
process of a platform:
\begin{enumerate}
    \item The \gls{SEC} is the initial code running, it switches from real 
mode to protected mode, initializes the memory space to run stack-based 
C code, and discovers, verifies and executes the next phase.
    \item The \gls{PEI} initializes permanent memory, handles the different 
states of the system (\eg recovery after suspending), executes the next phase.
    \item The \gls{DXE} discovers and executes drivers which initialize 
platform components.
    \item The \gls{BDS} chooses the boot loader to execute.
    \item The \gls{TLS} handles special applications or executes the boot loader from
 the \gls{OS}.
    \item The \gls{RT} is when the \gls{OS} executing, but 
there are still runtime services of firmware available to communicate with 
the \gls{OS}.
    \item The \gls{AL} takes control back over the \gls{OS} when it has 
shutdown.
\end{enumerate}

The \gls{TCG} specifies requirements for the measurement of \gls{UEFI}-compliant
firmware in \gls{TPM} \gls{PCR}s~\cite{tcg2014EFI}.
The \gls{SCRTM} in UEFI-compliant firmware is generally formed by the SEC
and PEI phases~\cite{zimmer2010beyond}, although no strict
requirements about its location are specified in the \gls{TCG} specification
as it can also be the entire BIOS~\cite{tcg2005client}.
Moreover, in recent platforms, the \gls{SCRTM} is a hardware-protected component, 
outside of the BIOS, that performs measurements on the BIOS before its
execution~\cite{intelBootGuard, hpSureStart}.
In our work, however, we only consider a non-hardware protected \gls{SCRTM}.

\section{Approach Overview}
\label{sec:overview}
BootKeeper is an offline analysis approach leveraging state-of-the-art binary analysis techniques to evaluate the validity and correctness of boot firmware images.

\subsection{Threat Model}
\label{sec:threat_model}
Our approach targets systems that implement measured boot protection mechanisms by using Trusted Computing technology.
From a high-level perspective, an attacker may attempt to tamper with a system's firmware in two ways:
\begin{itemize}
	\item By exploiting weaknesses of the \gls{SCRTM}, \eg a buggy or incorrect \gls{SCRTM} may only perform partial measurements. In this case, an attacker may be able to inject code within the vendor's firmware image in the non-measured portions of the memory.
	\item By directly injecting a malicious \gls{SCRTM}, the attacker may spoof the vendor's golden measurement values to pretend that the legitimate firmware is in place, while executing a malicious version of it.
\end{itemize}

By successfully circumventing the measurement process, an attacker may not only compromise the integrity of the system
while tricking the attestation procedure into reporting a legitimate software platform, but it may also leak secret
information from the \gls{TPM} such as cryptographic keys used for full disk encryption (as used by
Microsoft Windows's BitLocker, among other software products relying on this mechanism).

In the remainder of this paper, we assume the following attacker model:
\label{sec:attacker_model}
\begin{enumerate}
	\item The attacker does not have physical access to the system.
	\item The attacker does not have any form of privileged access to the system (neither local or remote, \ie no control over the \gls{OS}).
	\item The system itself has not been infected prior to the attack and is non-malicious (\ie the \gls{SCRTM} is invoked from a non-malicious environment).
	\item The \gls{SCRTM}'s code does not implement user input mechanisms (but such mechanisms may be implemented as part of later stages of the EFI boot process).
	\item The attacker has the ability and sufficient knowledge about the target platform to craft malicious firmware images, \ie access to the vendors' official firmware images, and knowledge about the platform's golden values (\ie correct measurement values), or the ability to obtain those by reverse engineering.
	\item The attacker may spread malicious images online (\eg by compromising the vendor's website or through third party websites such as user forums).
	\item Optionally, the attacker may remotely interfere with the automated firmware update process that comes with some systems by compromising the download site, or by mounting a man in the middle attack when applicable (\eg if this process does not check the validity of SSL certificates), to trick the remote system into applying a firmware update using an attacker-chosen malicious image.
\end{enumerate}

\subsubsection{Signature Verification}
In order to successfully install a malicious firmware image in the target system, an attacker needs to bypass eventual signature verification mechanisms in place. While this aspect is outside of the scope of this paper, we briefly demonstrate the practicality of this assumption as follows.

A good practice when releasing software updates is to rely on cryptographic signatures in order to guarantee the integrity of the new software image before or during the installation process.
Unfortunately, this process is often imperfect, as several vendors do not implement proper signature verification mechanisms, leaving gaps for an attacker to use a forged firmware image.
In other situations, attackers may use a stolen certificate~\cite{kim2017} to sign malicious firmware images, or may remotely exploit a vulnerability in the firmware update routine, to bypass the signature checks. In the remainder of this paper, we assume that the signature verification mechanism is either absent, or vulnerable.

\subsection{Analysis Steps}
Our analysis approach relies on the verification of a set of key properties, which we describe in more detail the remainder of this section.

\subsubsection{Code Integrity Properties (CIP)}
\label{sec:cip}

The \gls{SCRTM} code is always implemented
with two main fundamental operations: (1) an operation of fingerprinting, which scans the code regions in memory (\eg using SHA-1) and (2) a \gls{TPM} write operation, storing the computed fingerprints in the \gls{TPM}. 
We define these two operations as the building blocks of any \gls{SCRTM} measurement process performed on the platform. 

\begin{figure}[h]
    \centering
    \resizebox{\columnwidth}{!}{
        \tikzset{
    function/.style = {rounded corners, drop shadow, text centered, fill=white, draw},
    tpm/.style = {drop shadow, text centered, fill=white, draw},
}
\newcommand*\circled[1]{\tikz[baseline=(char.base)]{
        \node[shape=circle,draw,inner sep=0.5pt] (char) {#1};}}
\begin{tikzpicture}[-{Latex[length=0.6em]}, auto]
    \begin{scope}[local bounding box=Firmware]
        \node[function, align=center] (hash) {Legitimate\\Hash Function};
        \node[function, fill=gray!50, right=6em of hash, align=center] (legitimate) {Legitimate\\Firmware};
        \node[function, below right=5em of hash, xshift=-3em, align=center] (malicious) {Malicious\\SCRTM};
    \end{scope}
    \node[draw, fit={(Firmware)}, inner sep=0.9em, inner ysep=1.4em, yshift=0.3em, xshift=0.3em] (FirmwareBox) {};
    \node[below right] at (FirmwareBox.north west) {\textbf{BIOS}};
    
    \node[tpm, right=2em of FirmwareBox, yshift=-2.8em] (tpm) {TPM};
    
    \draw (malicious.north west) to [out=180, in=-100, looseness=1] node[below left, xshift=1em, align=center]
    {\circled{1} Calls\\with forged\\parameters} (hash.south);

    \draw (hash.south east) to [out=-20, in=60, looseness=1] node[right, xshift=1em, yshift=-1em, align=center] {\circled{3} Returns forged\\fingerprint} (malicious.north);
    \draw ([yshift=0.6em]malicious.east) -- node[below left, xshift=2.7em, align=center] {\circled{4} Sends forged\\fingerprint} ([yshift=0.6em]malicious.east -| tpm.west);
    \draw (hash) -- node[above] {\circled{2} Measures} (legitimate);

    \begin{customlegend}[legend cell align=left,
    legend style={at={(1.09,-3)}},
    legend entries={
        \footnotesize Executed,
        \footnotesize Not executed
    }]
        \addlegendimage{legend image code/.code={\node[draw, minimum width=2em, minimum height=0.8em] {};}}
        \addlegendimage{legend image code/.code={\node[draw, minimum width=2em, minimum height=0.8em, fill=gray!50] {};}}
    \end{customlegend}
\end{tikzpicture}
    }
    \Description{A figure describing a measurement-spoofing attack where the malicious BIOS sends a forged fingerprint to the TPM. The BIOS contains three blocks: a malicious SCRTM, a legitimate hash function, and a legitimate firmware. The malicious SCRTM calls the hash function with forged parameters to only measure the legitimate firmware which is not executed. Then, the malicious SCRTM sends the forged fingerprint to the TPM.}
    \caption{Example of a measurement-spoofing attack where an attacker sends a legitimate fingerprint of non-executed firmware.}
    \label{fig:forgery}
\end{figure}
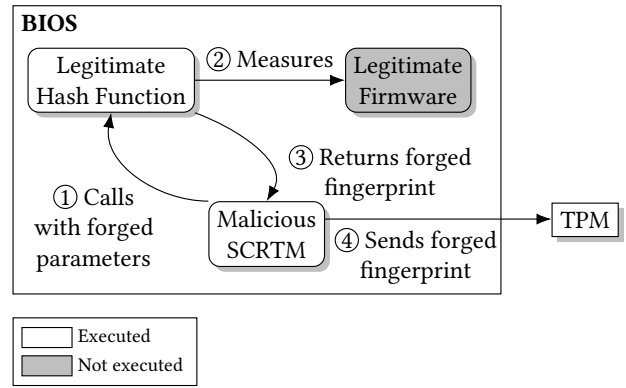

In order to elude the measurement process, an attacker may act at two different
levels. 
\begin{itemize}
    \item Firstly, as illustrated in~\autoref{fig:forgery}, the attacker may modify the fingerprint
          function (\eg code or parameters) to generate spoofed measurement values which
          correspond to valid golden values (\ie values corresponding to correct measurements on the
          vendor's firmware) even though the original firmware code is modified.

    \item Secondly, the attacker may modify the results of the fingerprint function just before
          these are written to the \gls{TPM}.
\end{itemize}
The \textit{tick} and the \textit{flea} attacks, described by \citet{butterworth2013bios},
are concrete examples of such attacks.

In order to prevent these attacks, 
our system verifies the three following properties:
\begin{enumerate}
\item {\it The authenticity of cryptographic hash functions}. Regardless of any potential implementation variants,
      our system must be able to verify the authenticity of the code used as part of the fingerprinting measurement
      process.
      BootKeeper leverages binary analysis techniques to verify that the correct hash function is indeed used
      as part of the firmware's fingerprinting code.

\item {\it The atomicity of the measurement process}. A correct \gls{SCRTM} implementation should also guarantee
      the atomicity of its measurement process, \ie that the fingerprinting and \gls{TPM} write operations are invoked
      sequentially in the correct order, and that the integrity of the measurement values if preserved between these two
      operations.
      In order to verify this property, BootKeeper constructs a \gls{CFG} of the \gls{SCRTM}, and detects eventual
      operations modifying the measurement results before those are written to
      the \gls{TPM}.\footnote{In practice, such operations may either correspond to malicious code attempting to forge measurement values, or to benign buggy code reporting erroneous measurements.}
\end{enumerate}

These two properties ensure the correctness of the \gls{SCRTM}'s code measurements process.
In addition to these, BootKeeper also ensures that the firmware under analysis does not present risks of certain classes of runtime attacks, as described below.

\subsubsection{Code Execution Integrity Property (CEIP)}
\label{sec:ceip}

Even if properties (1) and (2) are guaranteed, an attacker may attempt to alter the control-flow of the \gls{SCRTM} by redirecting the execution to malicious code hidden in the binary firmware image.
Fortunately, UEFI firmware runs in an execution environment protected by Data Execution Prevention (DEP). In other words, an attacker cannot trivially execute code injected in arbitrary sections of the binary image.

\textit{In the execution context of the \gls{SCRTM}, an attacker does not have the ability to inject code at runtime} since the SCRTM's code does not implement user input mechanisms and the \gls{SCRTM} code is invoked from a non-malicious environment (see rules 3 and 4 of our attacker model in~\autoref{sec:attacker_model}).
However, it remains possible for the attackers to hide code within the binary firmware image, and to attempt to trigger its execution at runtime.

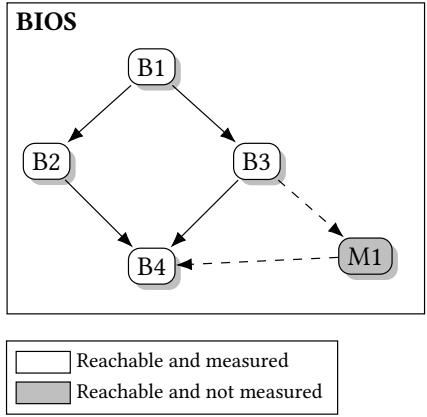
\begin{figure}[h]
    \centering
    \resizebox{0.7\columnwidth}{!}{
        \tikzset{
    block/.style = {rounded corners, drop shadow, text centered, fill=white, draw},
}
\begin{tikzpicture}[-{Latex[length=0.6em]}, auto]
    \begin{scope}[local bounding box=Firmware]
        \node[block] (B1) {B1};
        \node[block, below left=3em of B1] (B2) {B2};
        \node[block, below right=3em of B1] (B3) {B3};
        \node[block, below=6em of B1] (B4) {B4};
        \node[block, fill=gray!50, below right=3em of B3] (M1) {M1};
    \end{scope}
    \node[draw, fit={(Firmware)}, inner sep=0.9em, inner ysep=1.4em, yshift=0.3em, xshift=0.3em] (FirmwareBox) {};
    \node[below right] at (FirmwareBox.north west) {\textbf{BIOS}};

    \draw (B1) -- (B2);
    \draw (B1) -- (B3);
    \draw (B2) -- (B4);
    \draw (B3) -- (B4);
    \draw[dashed] (B3) -- (M1);
    \draw[dashed] (M1) -- (B4);

    \begin{customlegend}[legend cell align=left,
    legend style={at={(2.18,-3.2)}},
    legend entries={
        \footnotesize Reachable and measured,
        \footnotesize Reachable and not measured
    }]
        \addlegendimage{legend image code/.code={\node[draw, minimum width=2em, minimum height=0.8em] {};}}
        \addlegendimage{legend image code/.code={\node[draw, minimum width=2em, minimum height=0.8em, fill=gray!50] {};}}
    \end{customlegend}
\end{tikzpicture}
    }
    \Description{A figure with five basic blocks: four benign and one malicious. All are reachable, all the benign ones are measured, while the malicious one is reachable but not measured.}
    \caption{Example of an incomplete measurement of the firmware where reachable (malicious) code is not measured.}
    \label{fig:cfg}
\end{figure}

BootKeeper addresses this family of attacks as well by relying on an additional property:
\begin{enumerate}[resume]
    \item {\it Completeness of the measurements}.
          The \acrshort{SCRTM} must guarantee the completeness of the measurements of the firmware code memory regions.
	  More in detail, every memory region belonging to the \gls{CFG} of the \gls{SCRTM} must be measured and
          reported to the \gls{TPM} component.
	  By doing so, attempts to hide malicious code within non-measured memory regions is detected, as
	  represented in~\autoref{fig:cfg}, showing benign (B) and malicious (M) basic blocks forming a \gls{CFG}.
\end{enumerate}

We emphasize that BootKeeper does not rely on a-priori knowledge of the legitimate \gls{CFG} of firmware images.
Instead, the goal of BootKeeper is to ensure that \emph{all reachable code} will be \emph{correctly} measured and
reported to the \gls{TPM} at runtime.
As such, the detection of malicious code is a two-stage process: a static part (BootKeeper) which ensures that the
verification code is correctly implemented, and a dynamic part (the measured boot process), which relies on
those mechanisms.

Recall the attacker model presented in~\autoref{sec:threat_model}: (3) the \gls{SCRTM} executes in a non-malicious environment (\ie the initial state of the system is non-malicious, only firmware images are) and (4) it does not implement I/O mechanisms. As a consequence, dynamic code injection attacks are specifically excluded.

\section{System Design and Implementation}
\label{sec:approach}
In this section, we present the algorithmic properties of our analysis components.

\subsection{Code Integrity Properties (CIP) Validation}

The process of validating the Code Integrity Properties presented in~\autoref{sec:cip} relies on four analysis steps:
(1) Detecting \gls{TPM} write operations inside the firmware code,
(2) Detecting hash functions inside the firmware code,
(3) Validating the authenticity of hash functions,
(4) Validating the atomicity of the measurement process.
We now describe how the system achieves each step of the analysis.

\subsubsection{TPM Write Operation Detection}
The first step of our analysis is to identify the \gls{TPM} write
operations inside the firmware code. Such operations can be identified
by searching for standard API prototypes. According to the \gls{TPM}
specification version 1.2~\cite{tcg2005TIS, tcg2005client}, such 
functions must use a fixed address (\verb~0xFED40000~ is the default), 
along with an offset used for distinguishing among different
operations (\eg read and write) on the \gls{TPM} registers.

Unfortunately, we cannot predict how the compiler organizes the
instructions or how the code of the firmware performs the write
accesses. Developers tend to create abstraction layers (\eg to avoid
redundancy or to have a modular code), and may use different
optimization flags for the compilation of the firmware.
As a result of this, \gls{TPM} read and write operations do not straightforwardly appear as an offset from the specified constant address value in the
binary executable version of the firmware. Therefore, in order to tackle this problem and find a state of
the program where a write access to the \gls{TPM} known
address happens, we leverage symbolic execution, starting at
the entry point of the firmware, and record every instruction writing at the specified address
(\verb~0xFED40024~ in our case). This step of symbolic execution allows BootKeeper to resolve computed addresses for which it would be extremely difficult to reason about in a purely static setup. 
We leverage the angr~\cite{angr} platform to perform symbolic execution. 
During symbolic execution, the system tracks the state of registers and memory throughout program execution along with the constraints on those variables. Whenever a conditional branch is reached, the execution follows both paths while applying constraints to the program state to reflect whether the condition evaluates to True or False.  
At the end of this analysis step, BootKeeper has obtained a list of addresses in the firmware corresponding to the
\gls{TPM} write operations, if any such operation is present.
If the system does not find any \gls{TPM} write operation, it flags the firmware image as non-valid.

\subsubsection{Detecting Hash Functions}
\label{subsec:slicing}
The second step of our analysis is to identify hash functions
among the functions used by the firmware image to elaborate the measurement values.
 We apply the following algorithm. The analyzer starts from the
\gls{TPM} write operations and works iteratively on the
instructions flow in reverse order. To this end, we leverage a static backward approach~\cite{kiss2003interprocedural}. 
where for each identified \gls{TPM} write operation, our analyzer computes a backward slice starting from the
sensitive data. Sensitive data are the parameters of the
\gls{TPM} write operations, in our case the measurements value of
the hash function stored in a particular memory region. The aim of this 
backward slicing analysis step is to identify modifications on
sensitive data, and to find all the data sources from which the modified
data is derived.

The backward slice technique allows us to focus our analysis only on
the instructions that lead to a single \gls{TPM} write access, 
which is important for two reasons. First, for performance reasons, 
focusing on a subset of the program greatly improves the time needed to
perform further analyses. Second, and more importantly, it 
ensures that the \gls{TPM} write access is connected to a
measurement computed earlier in the execution (\ie not inserted in an ad-hoc manner).
Hence, given a set of instructions related to the \gls{TPM} write
access, BootKeeper detects if an actual measurement is present. 

The output of this analysis step is a set of traces corresponding to instructions correlated with the \gls{TPM} write
operations parameters.
Such traces could also be leveraged to locate the hash function code as well.
However, it is worth noting that the backward slicing algorithm returns an over-approximation of the instructions
leading to the program point where the \gls{TPM}
measurements are sent, and therefore, it might include unrelated instructions.
For example, functions which simply move computed hash values from one
memory structure to another may also be whitelisted by this analysis.
For this reason, we cannot simply rely on this technique to accurately locate the hash functions themselves,
and we leverage a different approach to precisely locate those, as presented in the following paragraph.

\subsubsection{Validating the Authenticity of Hash Functions}
\label{sec:hash}
\label{sec:sub:sub:discovery_sha1}
After BootKeeper has identified the set of instructions related to a particular
\gls{TPM} operation, it extracts the corresponding blocks and attempts to recognize one of the possible valid hash function (\eg SHA-1).
In order to automatically identify cryptographic functions within binary code, we leverage the approach presented by \citet{lestringant2015automated} which relies on \gls{DFG} isomorphism. From a high-level perspective, this approach compares the code structure of known functions with the code structure of unknown functions, to determine if these implement the same algorithm.
This approach first employs a normalization step (based on a code rewrite mechanism), which is designed to increase the detection capability by erasing the peculiarities of each
instance of an algorithm. Then, by relying on a sub-graph isomorphism algorithm, the normalized \gls{DFG} is compared to that of known reference functions. We chose this approach to recognize functions of well-known libraries which are used in real-world boot firmware images (including Crypto++ and OpenSSL~\cite{openssl}). 

BootKeeper leverages this technique starting from the instruction traces highlighted in the previous analysis step. From there, it creates a \gls{DFG} for each corresponding function of the firmware image involved in the trace, and attempts to match the signature of a standard hash function, such as the SHA-1 implementation of libOpenSSL.
The output of this analysis step is the set of basic blocks belonging to the identified hash function. If no known hash function is found, then \emph{BootKeeper flags the firmware image as non-valid.}

\subsubsection{Validating the Atomicity Property}
\label{sec:atomic}
\label{sec:sub:sub:operations-atomicity}
The final step involved in validating code integrity is to ensure
that no modification of the computed measurements occurs before those are written to the \gls{TPM}. 
From the instruction traces obtained during the backward slicing step (\autoref{subsec:slicing}),
BootKeeper  executes the corresponding code paths symbolically,
from the hash function's return instruction to the \gls{TPM} write operation, and rules out the presence of
instructions modifying the computed hash value.
In order to detect such instructions, BootKeeper performs a last step of forward reaching definition analysis on each
identified code path to ensure that the value stored in the \gls{TPM} indeed corresponds to the return value
of the correct hash function.
If any instruction on a given path modifies the measurement value, the atomicity property is violated.
In this case, the firmware image is reported as invalid, and BootKeeper reports the faulty instruction and program path.

\subsection{Code Execution Integrity Validation}
Recall that, in addition to verifying the correctness of the measurement process, BootKeeper also evaluates the risks of runtime attacks through a Code Execution Integrity Property (CEIP) described earlier in Section~\ref{sec:ceip}, which consists in the completeness of measurements.

Conceptually, if the measurement process is incomplete, \ie leaving out portions of the code section, it becomes possible for an attacker to ``hide'' malicious code in the non-fingerprinted areas, hence the importance of this property.

\subsubsection{Completeness of the Measurements}
\label{sec:coverage}

BootKeeper ensures that every function in the \gls{CFG} of the \gls{SCRTM} is measured. The acute reader may wonder what becomes of basic blocks of code which is deemed non-reachable by the control-flow recovery analysis: this point is discussed in~\autoref{sec:discussion}.
The \gls{CFG} is computed by a recursive algorithm which disassembles and analyzes each
basic block, identifies its possible exits (\ie successors) and adds them to the
graph. It repeats this analysis recursively until no new exits are
identified.
\gls{CFG} recovery has one fundamental challenge: indirect
jumps. Indirect jumps occur when the code transfers control flow to
a target represented by a value in a register or a memory location
(\eg \verb|jmp %eax|). Indirect jumps can be categorized in two main
classes: (1) Computed, an example could be an application that uses
values in a register or memory to determine an index into a jump table
stored in memory; (2) Runtime binding, \ie function pointers
which jump targets are determined at runtime.
BootKeeper leverages state-of-the-art analysis techniques for control-flow recovery, as available in the
\verb|angr| framework~\cite{angr}.
This process leverages a combination of forced execution, backwards slicing, and symbolic execution to recover,
to the extent possible, all jump targets of each indirect jump.
While it may not always be possible to recover all jump targets in complex software projects,
it is practical in the context of a boot firmware's \gls{SCRTM} due to the minimal aspect of such a code base.
A more detailed discussion of this point and the practical limitations that it may involve
is provided in~\autoref{sec:discussion}.

Once BootKeeper has obtained a \gls{CFG} of the \gls{SCRTM}, 
it proceeds to verify the coverage of the \gls{SCRTM} in terms of code fingerprinting.
In order to do so, BootKeeper analyses all input values passed to the fingerprinting functions
(\ie hash function identified in~\autoref{sec:hash}) to determine the address and size of the memory areas used to compute the measurements.
At this point, BootKeeper has statically identified the addresses and sizes of the memory regions that will be
fingerprinted by the \gls{SCRTM} at runtime. Its next step is to ensure that all reachable code in the SCRTM's
\gls{CFG} does indeed fall within the measured memory regions.
In order to do so, BootKeeper verifies that the address of each basic block belonging to the \gls{CFG} falls within that range. 
If it is not the case, it means that a part of the firmware's code will not be measured correctly, and \emph{flags the firmware image as non-valid.}

By ensuring that all the reachable code in the \gls{CFG} is indeed measured by the \gls{SCRTM}, BootKeeper prevents firmware modification attacks. where malicious code is inserted within the executable paths of the firmware's code.

\section{Experimental Evaluation}
\label{sec:evaluation}
This section presents our experimental results.
Our evaluation metrics cover multiple attack vectors representing a large span of the attack surface against
the \gls{SCRTM}.
These include the ability of our prototype to identify and recover the location of \gls{TPM} write instructions,
the effectiveness of our approach to detect the presence of forged \gls{TPM} measurements, and to detect possible hidden
code areas which are left unmeasured.
In addition to this, we also evaluate the robustness of our analysis against various compiler optimization settings.

\subsection{Experimental Setup}
We evaluate BootKeeper on two real-world implementations of boot firmware used in the industry, as well as a
custom-crafted malicious firmware implementing state-of-the-art attacks.
\begin{enumerate}
    \item \SeaBIOS~\cite{seabios}, a native x86 BIOS implementation with \gls{TPM} support.
          It also supports standard BIOS features and calling interfaces that are implemented by a typical proprietary
          x86 BIOS implementations.
          This project is meant to provide an improved and more easily extendable implementation in comparison to the
          proprietary counterparts which come as stock firmware on standard x86 hardware,
          and can be deployed as replacement firmware on a variety of motherboards.

    \item \EDKII~\cite{edk2}, a modern, cross-platform firmware platform supported by a number of real Intel and
          ARM hardware platforms.
          It is a component of Intel's TianoCore (Intel's reference implementation of \gls{UEFI}).
          Major vendors (\eg Apple, ARM, or HP) contribute to its development, and it serves as a basis for a number
          of proprietary UEFI-compliant firmware implementations.

    \item A custom-crafted malicious firmware image which we implemented to reproduce
          multiple variants of the state-of-the-art attacks introduced by \citet{butterworth2013bios}.
\end{enumerate}

For validating cryptographic
functions, we are using firmware (\EDKII) that implements the SHA-1
function of the OpenSSL~\cite{openssl} libraries and \verb~gcc~ as a compiler.

\subsection{TPM Write Access Detection}
We evaluated BootKeeper against \SeaBIOS, \EDKII, and our custom firmware.
Our hypothesis is that complex firmware implementations, like \SeaBIOS{} and \EDKII,
use abstraction layers which tend to obfuscate \gls{TPM} write instructions,
\ie these do not exhibit such instructions in the form of a fixed address corresponding to the specification
(\eg \verb~mov 0xfed40024, al~).
Additionally, these abstraction layers may invoke several functions to initialize data structures,
perform hardware tests, use loop structures or handle
different types of errors, thus creating many paths, potentially leading our analysis to path explosion during its symbolic execution phase.
In order to test the resiliency of our approach against the aforementioned intricacies, but also against compiler optimizations, we evaluated BootKeeper in the context of multiple compiler optimization settings. For each firmware implementation, we produced 5 variants using different optimization flags: \verb~-O1~ to \verb~-O3~ which optimize for speed, \verb~-Os~ which optimizes for size, and \verb~-O0~ for no optimization. This process generates 15 firmware variants.
We present the results in~\autoref{tbl:accuracy_find_writes}.

\begin{table}
    \centering
    \caption{Detection of the function writing the measurements}
    \label{tbl:accuracy_find_writes}
    \begin{tabular}{cccc}
        \toprule
        \specialcell{Optimization\\flags} &
        \specialcell{Custom\\firmware} &
        \EDKII &
        \SeaBIOS \\
        \midrule
        \verb|-O0| & \cmark & \xmark & \emark \\
        \verb|-O1| & \cmark & \xmark & \cmark \\
        \verb|-O2| & \cmark & \cmark & \cmark \\
        \verb|-O3| & \cmark & \cmark & \cmark \\
        \verb|-Os| & \cmark & \cmark & \cmark \\
        \bottomrule
    \end{tabular}

    \vspace{1em}
    \cmark~:~detected~\xmark~:~not~detected~\emark~:~error
\end{table}

BootKeeper can successfully detect the \gls{TPM} write access in 80\% of all cases (12 out of 15).
During this evaluation phase, we set an analysis timeout threshold of 10 minutes.

It is worth noting that BootKeeper found all the write access instructions within all the tested variants,
with the notable exception of \textit{unoptimized} \EDKII{} images.
We explain this observation by the fact that a lesser optimization level means more instructions to execute,
more loops and therefore more likelihood of path explosion during symbolic analysis.
While extending the analysis timeout threshold would be a viable option, we argue that vendors typically compile their
code with size optimizations (using \verb~-Os~) when releasing firmware for production use,
to reduce memory footprint,
in which case our approach succeeds under the 10 minutes threshold.

In the case of \SeaBIOS, the analysis results are missing in the situation where no optimization is used (\verb~-O0~).
Unfortunately, a bug in \SeaBIOS{} prevented the code from compiling with this optimization level at the time of our
evaluation, hence we could not test it.
Nonetheless, BootKeeper can successfully detect \gls{TPM} write operations in \SeaBIOS{} in all other optimization
settings.

In summary, these experiments demonstrate that our approach can correctly detect the \gls{TPM} write operations even
in intricate cases, in real-world firmware implementations, from legacy \gls{BIOS} to recent \gls{UEFI}-compliant firmware.

\subsection{Forged TPM Measurements Detection}
\label{sec:sub:forged-tpm-measurements-detection}
In order to evaluate the effectiveness of our approach against state-of-the-art attacks, we reproduced multiple variants of the \textit{tick} attack, described by \citet{butterworth2013bios}.
Each developed attack variant attempts to forge the measurements sent to the
\gls{TPM} by relying on several techniques: (1) attacking the SHA-1
function, (2) leaving out non-measured code, (3) modifying the \gls{TPM} parameters.
We now describe how BootKeeper performs against these attacks.

\paragraph{Hash Function Validation}
A first attack attempts to subvert the hash function results. In order to
test this attack, we create two firmware versions: one without a SHA-1
function, and another with a modified version of SHA-1 which
returns a tampered hash value. In either case, BootKeeper reports no 
match when generating signatures for the function defined inside the instruction 
traces (compared against a generated database of common cryptographic implementations from Crypto++ and libOpenSSL) and reports a violation of the property ``valid hash function''.

\paragraph{Completeness of the Measurement}
We evaluate this property by implementing an attack that can
modify the firmware's code 
by including
new code to the \gls{CFG} of the program. More specifically, we add new malicious
functions to a non-measured code area that is invoked before returning from the \gls{SCRTM} code.
Since this code is not included in any measurement performed by the hash function,
BootKeeper can detect it by ensuring that every part of the \gls{CFG} is correctly measured
(as described previously in~\autoref{sec:coverage}).

\paragraph{Atomicity Property}

In order to validate the atomicity property, we define a scenario
where an attacker properly measures the firmware with the correct SHA-1
function, but then tampers with the results by overwriting the measurements with other values before sending those to the \gls{TPM}, thus violating the atomicity property of the measurement operations.
During the analysis, BootKeeper can fetch the parameters of the hash function, and more importantly the
address of the pointer where the hash is stored.
Then, by performing the symbolic execution step described in~\autoref{sec:atomic}, it can detect the malicious instruction responsible for overwriting the measurements.

In order to fully reflect practical analysis challenges, we also consider the 
case where an attacker attempts to trick the analysis by incorporating a real measurement using the correct hash function, and later writing the measured value somewhere else before sending a forged value to the \gls{TPM}. We reproduce this attack scenario to evaluate the resilience of our analysis step presented in Section~\ref{sec:atomic} against the presence of false positives involved by the backward slicing algorithm.
In our evaluation, false positives indeed occur, but these are effectively filtered by the following step of forward reaching definition analysis (as expected).
In this situation again, BootKeeper can detect the attack, and report the malicious instruction responsible for overwriting the measurements.

\subsection{Performance}
While performance is not critical in the context of an offline analysis setting, we demonstrate the practicality of BootKeeper in terms of analysis time and memory usage.
In the following, we use a valid firmware for evaluation, and measure the required amount of time BootKeeper takes to proceed, while monitoring the peak of RAM usage. In this experiment, we use a single-thread on an Intel Xeon E312 with 64GB of RAM. The results presented in~\autoref{tbl:performance} were obtained by running this experiment 100 times, and represent the mean, minimum, maximum values and the standard deviation of both the runtime (in milliseconds) and the RAM usage (in
megabytes).
\begin{table}
    \centering
    \caption{Time and RAM usage for firmware analysis}
    \label{tbl:performance}
    \begin{tabular}{llllll}
        \toprule
        Type &      & Minimum & Maximum  & Mean    & \specialcell{Standard\\deviation} \\
        \midrule
        Time & (ms) & 103 950 & 108 760  & 105 704 & 849 \\
        RAM  & (MB) & 522.3   & 522.8    & 522.7   & 0.1 \\
        \bottomrule
    \end{tabular}
\end{table} 
Executing the entire analysis (\ie validating the firmware) takes on average 1 minute and 48 seconds. The memory usage peaks at 522 megabytes.

\section{Discussion}
\label{sec:discussion}
BootKeeper is a purely static approach relying on advanced binary
program techniques to analyze firmware images. As any static approach,
it comes with some challenges. In this section, we describe in more
detail the nature of these challenges, and how BootKeeper addresses
those. Finally, we point to some practical limits of our approach and
propose alternative research directions of interest.

\subsection{Theoretical Limitations}
In order to detect violations of the properties introduced in
Section~\ref{sec:overview}, BootKeeper relies on a combination of
state-of-the-art static analysis techniques, which together provide
the basis for implementing the verification algorithms presented in
Section~\ref{sec:approach}. These techniques, however, are subject to
theoretical limits, which prevents our approach from reasoning about
certain classes of properties in all possible situations.

BootKeeper relies in particular on:
\begin{itemize}
    \item Static \gls{CFG} recovery to determine the set of possible execution paths of a firmware image.
	      The \gls{CFG} obtained from binary analysis is neither sound nor complete (the general problem of deciding if
          an arbitrary path in a program is executable is undecidable~\cite{ramalingam1994undecidability}).

    \item Symbolic execution and constraint solving, to reason about the possible concrete values of memory
		  and registers at arbitrary points of the execution.
		  Symbolic execution is subject to the well-known state explosion problem due to its exponential growing
		  nature, and the general problem of constraint (SMT) solving is NP complete.

    \item Data-flow analysis, to isolate program paths involving measurement values, to generate program
		  slices in order to isolate the instructions affecting these values, and more generally, to detect faulty
		  operations.
		  Reasoning about data-flow at the binary level requires accurate models of data structure recovery,
		  and is subject to the pointer aliasing problem~\cite{ramalingam1994undecidability}.
\end{itemize}

Our approach inherits from these general limitations.
We discuss the practical impact of these theoretical limitations in~\autoref{sec:practical} below.

\subsection{Practical Impact}
\label{sec:practical}
The following is a discussion of the practical impact of the aforementioned limitations in our approach. 

\subsubsection{False Positives}
In the case where the \gls{CFG} is too conservative and includes an overestimate of possible code paths in the graph,
BootKeeper will accordingly operate conservatively during the verification of property 3, \ie the completeness
of measurements.
While this may lead to false positives in certain circumstances, we stress that the \gls{CFG} we obtain from a binary has
a basic-block level granularity.
In comparison, vendors typically scan entire memory regions corresponding to sections from the firmware binary.
\textit{Why not just measure entire sections then?}
Our approach is more fine-grained, and aims to ensure that the vendor conforms to at least a defined minimal code
coverage corresponding to (an estimation) of the possible execution paths. 

In the case of an incomplete \gls{CFG} between the return value of the hash function used in the measurements and the
subsequent \gls{TPM} write operation,
BootKeeper will not be able to compute a backward slice and therefore will not be able to validate property 2,
\ie the atomicity of the measurement process.
In this case, \emph{it will flag the image as malicious}, thus generating a false positive.

\subsubsection{False Negatives}
Similarly, the \gls{CFG} may miss edges in the graph corresponding, such as indirect jumps caused by complex instances
of runtime binding which cannot be resolved even with symbolic execution.
This may happen, among other possible cases, when external information (\ie external to the program) is required to compute the jump target.
The presence of such obfuscated or evasive code in early stages of firmware execution is, by itself, an excellent indicator of maliciousness which our approach could be extended with.

When such constructs are benign and part of the official vendor's firmware, an attacker may 
succeed in hiding a payload $P$ if: 1) the \gls{SCRTM} omits one portion of executable memory $M_1$ during the measurements, 2) BootKeeper is missing a part of the \gls{CFG} corresponding to a set of basic blocks mapped in memory during runtime as $M_2$ (in practice, $M_2$ may not be contiguous), and the following holds true:
$$M1 \cap M2 \neq \emptyset \wedge |M_1 \cap M_2| \geq sizeof(P)$$
Without ruling out the possibility of strong attacker specifically challenging state-of-the-art static analysis techniques, we estimate that BootKeeper significantly raises the bar for an attacker to circumvent the measurement process, and we consider our approach practical in the context of a large span of possible attacks. 

The presented limitations are intrinsic to any static approach, and cannot trivially be addressed without additional knowledge of the runtime environment.  In the next section, we discuss possible alternatives to overcome these limitations.   
 
\subsection{Alternative Solutions}  
In order to analyze the program paths involved in firmware during execution in
a dynamic setting, an emulation of all the hardware components involved during
the platform initialization process would be necessary. Implementing such a
system is a cumbersome engineering task, especially if numerous targets need to
be supported.  An alternative approach is to directly instrument the hardware
to dump the state of registers and memory as the firmware executes, the
knowledge of which would ease offline analysis. Similar to this is the Avatar
approach~\cite{avatar} which selectively switches between different execution
models, in a setup which is backed by the physical hardware.

While relevant to this discussion, neither of these approaches fits
within the scope of this paper.  In comparison, BootKeeper requires no
hardware nor custom hardware models.

\subsection{Obfuscation}
Static binary program analysis techniques are vulnerable to the presence of obfuscation, and it is possible that a malicious firmware author could attempt to attack BootKeeper in this manner.
For instance, \citet{sharifimpeding} obfuscate conditional code by using the result of a hash function as a condition replacement. Since cryptographic hash functions have the pre-image resistance property, it is impossible for constraints solvers to solve all the constraints generated by the operations of the hash function.

These weaknesses are inherent to any tool relying on static program analysis~\cite{shoshitaishvili2015firmalice,chipounov2012s2e,yadegari2015symbolic}.

In the context of boot firmware, the problem related to obfuscation is two-fold. First, genuine vendors could use obfuscation techniques to protect their code against reverse engineering. Secondly, by relying on obfuscation techniques, an attacker could attempt to defend against automated program analysis. While the former would affect BootKeeper, the latter may be used as an indicator malice.

%%% Local Variables: 
%%% mode: latex
%%% TeX-master: "../paper"
%%% End: 

\section{Related Work}
To the best of our knowledge, John Heasman developed the first public
\gls{BIOS} rootkit by modifying \gls{ACPI} tables stored in
the \gls{BIOS}~\cite{heasman2006implementingACPI}, and he also
showed how to make a persistent rootkit by re-flashing the expansion
\gls{ROM} of a \gls{PCI}
device~\cite{heasman2007implementingPCI}. Other attacks have been
performed since then, Anibal Sacco and Alfredo Ortega discussed how to
inject malicious code in Phoenix Award
\gls{BIOS}~\cite{sacco2009persistent} and Jonathan Brossard
showed the practicability of infecting different kinds of
firmware~\cite{brossard2012hardware}. In addition to papers and proof
of concepts of attacks, some malware is also taking advantage of the
lack of security of the boot firmware. For example, the Chernobyl
virus~\cite{cih}, which appeared in 1999, tried to overwrite the
\gls{BIOS} to make it unbootable. In 2011, the malware
called Mebromi~\cite{mebromi} re-flashed the \gls{BIOS} of its
victims to later write a malicious \gls{MBR} which infected the
\gls{OS} even when it was re-installed
from scratch.

All these attacks can be detected if the vendor is trustworthy, a
\gls{TPM} device is present and used correctly.
Several misconfiguration and design issues, however, show that the \gls{TPM} can be
attacked as well.
In this direction, \citet{butterworth2013bios}
demonstrated a replay-attack that forges the
measurement sent to the \gls{TPM} to fake an uncorrupted
\gls{BIOS} in case of non-respect of the specifications and
recommendations.
\citet{bruschi2005replay} also showed a replay-attack in an
authorization protocol of the \gls{TPM}.
\citet{sadeghi2006tcg} and \citet{butterworth2013bios} revealed that some \gls{TPM}
implementations do not meet the \gls{TCG} specifications which
may have critical security implications.
\citet{kauer2007oslo} also
demonstrated a \gls{TPM} reset attack which
allows an attacker to forge the \gls{PCR} values.

Several approaches have been proposed
to improve the \gls{TPM} technology and the boot firmware integrity techniques.
For example, Bernhard Kauer proposed a counter
measure~\cite{kauer2007oslo} to the reset attack on the \gls{TPM}
by using a \gls{DRTM}. In the direction of firmware security, 
dynamic analysis using symbolic execution has been extensively used to
find vulnerabilities in
firmware~\cite{davidson2013fie,bazhaniuk2015symbolic,shoshitaishvili2015firmalice,zaddach2014avatar,kuznetsov2010testing}.
More related to our work, \citet{bazhaniuk2015symbolic}
used an approach to detect vulnerabilities in boot firmware. Our work
is orthogonal to such approach and focuses on boot firmware phases
where vulnerabilities are not detected or fixed by the vendor and they
can be used by an attacker to tamper with the boot process (\eg
to forge \glspl{PCR} values).

\citet{butterworth2013bios} designed a timing-based attestation at the
\gls{BIOS} level as an alternative to
the hashing of the firmware. Such a technique provides a reliable way
to attest the integrity of a platform even if the attacker has the
same privilege level as the \gls{SCRTM}. The idea, adapted from
previous work on timing-based
attestation~\cite{schellekens2008remote}, is that in the absence of an
attacker the time required to perform a checksum of the firmware will
be constant. When an attacker tries to fake the checksum, she requires
additional instructions that increase the execution time, hence it
can be detected by the system.
While this work greatly improves the trust in the remote attestation,
and fixes the vulnerabilities discovered in their paper, it requires a
complicated architecture for being deployed. In fact, it needs to set
up a remote server for the attestation phase and to modify the interrupt
signal handling in the \gls{OS} to obtain a precise measurement of
the code execution.
On the contrary, our approach works without having an attestation
architecture and it only performs static checks on the firmware boot
image.

Recent platforms incorporate immutable, hardware protected \glspl{SCRTM}, 
called Intel Boot Guard~\cite{intelBootGuard} and HP Sure
Start~\cite{hpSureStart}. They are immutable \glspl{SCRTM} that measure and
verify at boot time the BIOS before its execution, thus providing firmware
integrity and a trusted boot chain with a Root-of-Trust locked into
hardware. Such technologies ensure that the first measurement cannot
be forged, since the attacker cannot modify their code. Both
technologies, however, are only available in recent Intel and HP platforms.
In addition, Intel Boot Guard has been showed to be vulnerable to a certain
class of attacks~\cite{alex}.
The advantage of our approach with
respect to those new technologies is twofold. First, it can be used to
protect architectures that are not equipped with such hardware
features. Second, our approach is orthogonal to such hardware
protections, since BootKeeper can be used as a standalone analyzer from
the vendor side for validating the \gls{SCRTM} code as the last step of the
deployment process. The main contribution of BootKeeper is related to
the software properties that we devise for validating the measurement
process. When BootKeeper is used by the vendor, our analyzer can
perform the same analysis (\eg enforcing software properties) at the
source code level, and verify that no one tampers with the measurement
task during the developing process.

\section{Conclusion}
In this paper, we introduce BootKeeper, a binary analysis approach to validate the measurement process of a boot firmware.
Our system uses static analysis and symbolic execution to validate a set of software properties on the measurement process implemented as part of the UEFI \emph{measured boot} specification. 
BootKeeper detects incorrect implementations of UEFI firmware which do not exhaustively or correctly implement the measured boot process, as well as malicious images crafted with the intention of bypassing the measured boot process. More specifically, BootKeeper focuses on the \gls{SCRTM}, which is the most critical component in the verification chain.
An incomplete \gls{SCRTM} implementation leaves room for an attacker to hide code in subsequent parts of the firmware, whereas a malicious \gls{SCRTM} voluntarily ignores specific regions where malicious payloads are hidden, or attempts to forge measurements in order to match the measured values of a legitimate vendor firmware (\ie golden values), among other possible attacks.

This approach 
can greatly improve trust in boot firmware update procedures. We evaluate BootKeeper against real-world firmware used in the industry as well as custom malicious firmware images, and show that our system is able to detect multiple variants of a variety of attacks from the state-of-the-art in the literature.

\bibliographystyle{ACM-Reference-Format}
\bibliography{paper}

\end{document}